\documentclass[12pt]{article}
\usepackage{amsmath,amsfonts, amssymb, braket, bbold}
\usepackage[normalem]{ulem}
\usepackage{tikz}
\usetikzlibrary{trees,er,snakes,shapes,mindmap}
\usepackage{titlesec}
\titleformat{\section}
 {\normalfont\fontfamily{put}\fontsize{13pt}{16pt}\bfseries\color{black}}
{\thesection}{1em}{}
\titleformat{\subsection}
 {\normalfont\fontfamily{put}\fontsize{12pt}{16pt}\bfseries\color{black}}
{\thesubsection}{1em}{}
\def \beq  {\begin{equation}}
\def \eeq  {\end{equation}}
\def \beqar {\begin{eqnarray}}
\def \eeqar {\end{eqnarray}}
\allowdisplaybreaks
\def\sqr#1#2{{\vcenter{\vbox{\hrule height.#2pt
\hbox{\vrule width.#2pt height#1pt \kern#1pt
\vrule width.#2pt}\hrule height.#2pt}}}}

\def\Tr {{\rm Tr}}

\def\bD {\bar{D}}

\def\del {\partial}

\def\a {\alpha}
\def\b {\beta}
\def\e {\epsilon}

\def\C {{\cal C}}

\def\F {{\cal F}}

\def\I {{\cal I}}

\def\half{\textstyle{1\over 2}}

\mathchardef\mhyphen="2D
\begin{document}
\fontfamily{bch}\fontsize{12pt}{16pt}\selectfont
\def \CMP {{Commun. Math. Phys.}}
\def \PRL {{Phys. Rev. Lett.}}
\def \PL {{Phys. Lett.}}
\def \NPBProc {{Nucl. Phys. B (Proc. Suppl.)}}
\def \NP {{Nucl. Phys.}}
\def \RMP {{Rev. Mod. Phys.}}
\def \JGP {{J. Geom. Phys.}}
\def \CQG {{Class. Quant. Grav.}}
\def \MPL {{Mod. Phys. Lett.}}
\def \IJMP {{ Int. J. Mod. Phys.}}
\def \JHEP {{JHEP}}
\def \PR {{Phys. Rev.}}
\def \JMP {{J. Math. Phys.}}
\def \GRG{{Gen. Rel. Grav.}}
\begin{titlepage}
\null\vspace{-62pt} \pagestyle{empty}
\begin{center}
\vspace{1.3truein} {\large\bfseries
Fractional quantum Hall effect in higher dimensions}
\vskip .1in
{\Large\bfseries ~}\\
\vskip .1in
{\sc Abhishek Agarwal$^a$, Dimitra Karabali$^{b,d}$, V.P. Nair$^{c, d}$}\\
\vskip .2in
{\sl $^a$\,Physical Review Letters, 
American Physical Society\\ 
Hauppauge, NY 11788}\\
\vskip.1in
{\sl $^b$Physics and Astronomy Department,
Lehman College, CUNY\\
Bronx, NY 10468}\\
\vskip.1in
{\sl $^c$Physics Department,
City College of New York, CUNY\\
New York, NY 10031}\\
\vskip.1in
{\sl $^d$The Graduate Center, CUNY\\
New York, NY 10016}\\
 \vskip .1in
\begin{tabular}{r l}
{\sl E-mail}:&\!\!\!{\fontfamily{cmtt}\fontsize{11pt}{15pt}\selectfont 
abhishek@aps.org}\\
&\!\!\!{\fontfamily{cmtt}\fontsize{11pt}{15pt}\selectfont 
dimitra.karabali@lehman.cuny.edu}\\
&\!\!\!{\fontfamily{cmtt}\fontsize{11pt}{15pt}\selectfont vpnair@ccny.cuny.edu}\\
\end{tabular}
\vskip .5in

\centerline{\large\bf Abstract}
\end{center}
Generalizing from previous work on the integer quantum Hall effect, we construct the effective action for the analog of
Laughlin states for the fractional quantum Hall effect in higher dimensions.
The formalism is a generalization of the parton picture used
in two spatial dimensions, the crucial ingredient being the cancellation of
anomalies for the gauge fields binding the partons together.
Some subtleties which exist even in two dimensions are pointed out.
The effective action is obtained from a combination of the
Dolbeault and Dirac index theorems. We also present
expressions for some transport coefficients such as
Hall conductivity and Hall viscosity for the fractional states.

\end{titlepage}
\fontfamily{bch}\fontsize{12pt}{17pt}\selectfont
\pagestyle{plain} \setcounter{page}{2}
\section{Introduction}
The phenomenon of the quantum Hall effect (QHE) hardly needs any stress on its importance as it has been the topic of intense investigations, both theoretically and experimentally, over the last several decades
\cite{QHE-general}.
While most of the research has focused on two dimensions, 
already several years ago,
the enticing mathematical structure of QHE prompted
suggestions on generalizations to higher dimensions.
Even though these seemed to be mathematical curiosities
initially, it is interesting that QHE in higher dimensions
may in fact be experimentally realizable
using the idea of synthetic dimensions \cite{{4DQHE}, {4DQHE1}}.
The initial proposal for higher dimensional QHE considered
space as a 4-sphere $S^4$ \cite{ZH}.
Shortly after this, QHE on complex manifolds of arbitrary dimensions
were analyzed.
The explicit solution of the Landau problem, the construction
of integer quantum Hall states, the analysis of the
edge excitations, etc. were carried out leading to a uniform extension to
all higher even dimensions \cite{KN1}-\cite{KN4}, see also
\cite{{alexios}, {everyone}}.
A specific case of odd dimensions was also
investigated \cite{VPN-Rand}.
To a large extent, the problem is defined by topological considerations.
Since the lowest Landau level obeys a certain holomorphicity condition,
the Dolbeault index theorem can be used to analyze many features of the phenomenon \cite{EGH}.
It is then possible to show that a
Chern-Simons action associated to the Dolbeault index density
describes the bulk dynamics of a QHE droplet of fermions (for integer filling fractions),
including fluctuations of gauge and gravitational fields, as well as 
Abelian and nonabelian background magnetic fields
\cite{KN5}.
Needless to say, the specialization of this general effective action
to 2+1 dimensions agrees with explicit derivations 
based on wave functions carried out by many authors
\cite{WZ}-\cite{AG2}.
Higher dimensions also allow for an enlarged set of
transport coefficients. Some of these were recently worked out
in \cite{KN6}, where it was also explained how the band structures
of the electrons could be incorporated in the effective action.

All the higher dimensional generalizations considered so far have been for the integer QHE. In 2+1 dimensions, we also have a wealth of information
regarding the fractional QHE, including the wave functions
for many of the experimentally realized states
(such as the Laughlin, Jain and Moore-Read states), effective actions for the bulk and boundary dynamics of a droplet of fermions, various transport coefficients, etc.
In this paper, we consider the question of how quantum Hall states can be defined in higher dimensions for fractional filling,  in particular higher dimensional analogs of the
$\nu = 1/m$ Laughlin states in 2+1 dimensions, where $m$ is an odd integer. This is a natural next step for QHE and can be particularly relevant in the light of potential experimental realizations
 in higher dimensions. In 2+1 dimensions, a variety of methods exist to map the fundamental excitations, namely electrons, to composite particles whose known states map to fractional excitations for the electrons. One standard approach is that of flux attachment; see \cite{books} for pedagogical introductions to the topic of flux attachment, and \cite{Gromov-thesis} and \cite{AA-FQH} for reviews of aspects of flux attachment applied to various contemporary issues. A complementary approach available in 2+1 dimensions is that of the parton construction of the fundamental electrons \cite{partons1, partons2}.

While flux attachment is very natural in two spatial dimensions, so far we have not been able to find a workable extension to higher dimensions. Therefore in this paper we will consider the generalization of the parton picture. In this scenario, the fundamental fermion, i.e., the electron is viewed as a composite particle made of $m$ partons, one parton each of $m$ 
species. 
If $\psi$ denotes the electron field, and $q_i$, $i=1, 2, \cdots, m$, denote the parton fields, this is equivalent to the statement
\beq
\psi \sim q_1 \, q_2 \cdots q_m
\label{fqhe1}
\eeq
The partons themselves are taken to be fermions of charge $e/m$ 
and $m$ is taken to
be an odd positive integer, i.e., $m = 2 l +1$, $l= 0, 1, \cdots$, to obtain fermionic
statistics for the composite particle. (We hasten to add that
the partons themselves are not viewed as physical entities, rather this 
construction is a convenient mathematical device that encodes some
of the multiparticle strong coupling effects, in a way that is not yet fully understood.) 
One then constructs a state where the partons, in the external magnetic field, form an integer quantum Hall state, say, with filling fraction equal to $1$,
i.e., $\nu =1$,  for simplicity.
In terms of the electrons this state may be viewed as a state of filling fraction
$\nu = 1/m$.
For this strategy to be consistent, the occupation number in the given
$\nu =1$ state for each species of partons should be the same, i.e.,
$n_1 = n_2 = \cdots = n_m \equiv n$, so that we have $n$ electrons,
with one parton of each kind for each electron.
The equality of the $n_i$'s is enforced by use of a set of $U(1)$ gauge fields, which we will refer to as the $b$-fields in this paper.
The same fields can also be taken to be the agency binding the partons to form the electron as in (\ref{fqhe1}). 

The use of the $b$-fields to bind the partons and the use of the equations of motion for time-components of the $b$-fields to obtain the equality of the
$n_i$'s show clearly that they are dynamical fields, unlike the external
magnetic field or geometrical characteristics such as the metric and spin connection of the manifold.
In a functional integral approach, they are therefore to be integrated out.
This can be formally carried out in 2+1 dimensions, where, 
because the action is a Chern-Simons three-form and quadratic
in the abelian $b$-fields,
the integration can be done in closed form, leading to a framing anomaly \cite{framing}.
In the resulting effective action, this is equivalent to a gravitational Chern-Simons term, and hence to an additional gravitational anomaly for the edge modes of a quantum Hall droplet.
The generalization of the framing anomaly calculation to higher dimensions is not straightforward, but, as we shall now argue,
 there is an alternate way of viewing the
procedure of integrating out the $b$-fields. 

If we consider a quantum Hall droplet, the bulk Chern-Simons action is not gauge-invariant. The complete effective action is
gauge-invariant because there are edge modes on the boundary
of the droplet which cancel the variation of the
bulk terms under gauge transformation of the electromagnetic field
$A$
and the spin connection $\omega$.
In the case of integer QHE this cancellation helps to identify the nature of the edge excitations.
When we consider partons coupled to the $b$-fields, again, the bulk action is not gauge-invariant. This will lead to nonzero terms
on the edge under the gauge transformation of the $b$-fields as well.
While physical edge excitations, as before, can cancel the terms 
due to the gauge variation of $A,\, \omega$, 
physical excitations should not carry $b$-charges since the
partons and the $b$-fields constitute only a theoretical trick to
incorporate certain nonperturbative effects.
Therefore, we
consider a set of auxiliary fields, which we will refer to as the
spectator fields (using the terminology of 't Hooft),
which have an anomaly on the edge that
can cancel the variation of the parton bulk action 
under the gauge transformation of the $b$-fields.
(As we shall see shortly, the spectator fields will be chiral spinors.)
With this ``anomaly cancellation'', the dynamics of $b$-fields
is made consistent with no left-over CS-type terms for the $b$-fields
and one can integrate out the $b$-fields without worrying about a framing anomaly. However, the spectator
fields can contribute to the gravitational anomaly and this indeed captures
the effect of framing anomaly calculations.
This strategy of anomaly cancellation provides
a way to bypass the intricacies of deriving the framing anomaly
even for 2+1 dimensions.
Further, it can be generalized to
higher dimensions.

We emphasize that the spectator fields are defined on the boundary.
But we can associate a bulk action with the spectators, since
their anomaly (which is on the edge), 
via the standard descent procedure, can be
expressed as the gauge variation of a bulk Chern-Simons term.
This bulk term is a convenient way to encode the
anomaly of the spectator fields.
The bulk action due to the partons plus the bulk term associated with
the spectator fields
will then give the bulk effective action for the fractional quantum Hall state of interest,
but the cancellation of anomalies is really implemented
on the boundary of the droplet.

Our strategy for generalization of fractional QHE to higher dimensions will thus be as follows. We will consider the electron or the fundamental fermion to be made of $m$ partons. These partons will be coupled to a set
of $U(1)$ $b$-fields. The parton fields obey a holomorphicity condition
since we restrict them to a particular Landau level.
Therefore we can use
the Dolbeault index theorem to obtain the
bulk action of the partons,
along the lines of our earlier work \cite{KN5}.

The spectator fields do not couple to the electromagnetic field and
so do not fall into Landau levels.
Therefore, the anomaly for such fields must arise as in standard
field theory.
We will take the spectators to be chiral spinor fields as they are
the ones which can generate a gauge anomaly.\footnote{A chiral boson in 2 dimensions can generate an anomaly, but, via fermionization, this is
equivalent to a chiral spinor. A self-dual field in 4k+2 dimensions can generate gravitational anomalies but this is not relevant for the present paper since we consider 2+1 and 4+1 dimensional cases only \cite{anom}.
In two dimensions, a self-dual field is again equivalent to a chiral boson.}
The anomaly due to the spectators can then be calculated 
using the Dirac index theorem.

We emphasize that there is a distinction in the use of the index theorem for the partons and the spectators. The Dolbeault index theorem leads to the bulk action for the partons because they obey a holomorphicity condition. The Dirac index theorem for the spectators is used to calculate a gauge anomaly.

Once the anomaly cancellation has been ensured, the remaining
Chern-Simons terms, involving the external magnetic field and the spin connections, can be assembled to give the effective action for the
fractional quantum Hall state. This is basically made of the left-over terms in the
parton bulk action plus the bulk action which encodes the anomaly due to
the spectators.
Transport coefficients such as the Hall conductivity, the Hall viscosity, etc.
can then be read off this effective action.

In the next section we will give a brief 
summary of how the Dolbeault index theorem can
be used to derive the bulk effective action for integer QHE. 
This is to help set the stage
for the subsequent analysis.
We also indicate how the anomaly due to the spectator fields can be
identified.
In section 3 we will
consider the Laughlin states (of filling fraction
$\nu = 1/m$) in 2+1 dimensions.
Primarily this is to show how our modified parton picture recovers known results in 2+1 dimensions. However, it will also highlight some subtleties
in the application of the parton picture, even in 2+1 dimensions,
to states with other values for the filling fraction.
Further it will lay out the necessary mathematical steps for generalization to higher dimensions and the subtleties to be aware of.
In section 4 we will consider the construction of Laughlin-type states
in 4+1 dimensions. Anomaly cancellation for the dynamical
$b$-fields will again be the guiding principle.
The effective action, valid for generic four-dimensional spatial manifolds, is worked out, see (\ref{4d-11}), (\ref{122}).
The electromagnetic current and the energy-momentum tensor can
then be read off, they are given in (\ref{4d-12}), (\ref{140a}), (\ref{138a}).
It is then straightforward to write down the Hall conductivity and Hall viscosity, which are the primary response functions of interest. A key point worthy of remark is that
the leading term of the Hall conductivity has a factor $1/m^2$
in four dimensions, rather than $1/m$ as in two dimensions. We analyze fractional quantum Hall effect (FQHE) on the complex manifold $S^2 \times S^2$, as a special case of our general result, and comment on the dimensional reduction of the (4+1) dimensional results to 2+1 dimensions.
Section 5 gives a brief recapitulation of the key steps in our analysis
as well as the key results. We conclude with a number of remarks
on possible extensions of the present results.

Before we embark on the details of our construction, we note that there have been other attempts at generalizing the fractional quantum Hall effect
to higher dimensions \cite{heck}. The effective action in 2+1 dimensions can be obtained from a set of
Chern-Simons actions involving one-form fields, put together 
using a so-called $K$-matrix. Elimination of some of the fields via their equations of motion leads to the effective action in terms of the electromagnetic field.
In higher dimensions, a similar possibility is to consider Chern-Simons-like actions again, but using higher forms as the basic fields.
This can then lead to an effective action for fractional QHE. 
However, in this approach the basic constituents will be extended objects of suitable dimension coupling to the higher form fields, while, in our case, the basic constituents are particles coupling to the usual electromagnetic field.
Thus the two generalizations are {\it a priori} different; it would indeed be interesting to see if there is any way to relate them.

\section{The effective action from an index theorem }
In this section we give a brief resume of the derivation of
the bulk effective action from the index theorem
for integer QHE 
in arbitrary even spatial dimensions \cite{KN5}.
We consider the case where the spatial manifold
is a complex manifold and thus we will be using the Dolbeault index theorem
which is the relevant one for such cases.
However, once we obtain the effective action, it can be extended to
include small but arbitrary perturbations of the metric and spin connection 
which do not necessarily preserve the complex structure.
This will be relevant for deriving the Hall viscosity, for example.

The single particle Hamiltonian is of the form $-{\half} D_i \bD_{\bar i}$,
where $D_i$ and $\bD_{\bar i}$ are the holomorphic and antiholomorphic covariant derivatives. 
These include both gauge and gravitational fields; for the present
work, the gauge fields will include the $b$-fields
mentioned in the introduction as well as 
the external electromagnetic vector potential.
Ignoring the spin of the fermion (which can be the
electron or the parton), its wave function
is a complex function of the coordinates.
For the lowest Landau level the wave functions
obey the holomorphicity
condition
\beq
\bD_{\bar i} \Phi = 0
\label{fqhe2}
\eeq
The number of normalizable solutions to this equation is given by the index theorem for the twisted Dolbeault complex
as
\beq
{\rm Index}( \bD ) = \int_M {\cal I} =  \int_M {\rm td}(T_cM) \wedge {\rm ch} (V)
\label{fqhe3}
\eeq
where ${\rm td}(T_cM)$ is the Todd class on the complex tangent space of the spatial manifold $M$
and ${\rm ch}(V)$ is the Chern character of the relevant
vector bundle \cite{EGH}. 
The formula (\ref{fqhe3}) is rather cryptic, so we will give a brief explanation of
how it can be used to construct the effective action.

For a real manifold of dimension equal to $2k$, the holonomy group, which is the group corresponding to parallel transport, is $SO(2k)$.
This means that the spin
connections and curvatures take values in the Lie algebra of 
$SO(2k)$. However, for a complex manifold, we only allow
coordinate transformations which preserve the complex structure, so that
the notion of holomorphicity is preserved.
This restricts the
holonomy group to $U(k) \subset SO(2k)$.
The frame fields, viewed as one-forms, separate into
holomorphic forms and antiholomorphic forms.
These are combinations of the real ones
given by the complex structure. 
A corresponding set of combinations can be made for the tangent
space, which is, after all, dual to the forms. This leads to
$T_cM$. The Todd class in (\ref{fqhe3})
is given in terms of the curvature two-form for $T_c M$.
For any vector bundle with curvature ${\cal F}$, the Chern classes are defined by\footnote{We start with connections and curvatures in an antihermitian basis since they are natural
allowing us to write $F = d A + A A$, etc. This leads to some factors
of $i$ in various expressions at this stage,
since the hermitian fields are $i\F$, $iR$.
Later we will move to a hermitian basis.}
\beq
\det \left( 1 + {i \, {\cal F} \over 2 \pi} \,t\right) = \sum_i c_i \, t^i
\label{fqhe4}
\eeq One can then give an expansion of the Todd class in terms
of the Chern classes as \cite{EGH}
\beq
{\rm td} = 1 + {1\over 2} \, c_1 +{1\over 12} ( c_1^2 + c_2) + {1\over 24} c_1\, c_2
+ {1\over 720} ( - c_4 + c_1\,c_3 + 3 \, c_2^2 + 4\, c_1^2 \,c_2 - c_1^4) + \cdots
\label{fqhe5}
\eeq
A general expression for the Todd class valid for all dimensions
(and hence including higher rank
differential forms) is given in \cite{{EGH},{anom}}.
It is best expressed in terms of what is known as the
splitting principle. We do not quote it here, since, for the present purpose, 
equation (\ref{fqhe5}) will suffice.

The field $\F$ in (\ref{fqhe4}) is given by the curvature two-form $R$ for $T_cK$.
Explicit formulae for the first few Chern classes are then as follows.
\beqar
c_1 (T_c K)& = & \Tr ~{iR \over 2\pi} \nonumber \\
c_2 (T_c K)& = & {1 \over 2} \Biggl[ \Bigl(\Tr {iR \over 2\pi}\Bigr)^2 - \Tr \Bigl({iR \over 2\pi}\Bigr)^2 \Biggr] \nonumber \\
c_3 (T_c K) &=& { 1 \over 3!} \Biggl[ \Bigl(\Tr {iR \over 2\pi}\Bigr)^3 - 3\, \Tr {iR \over 2\pi} \,\Tr \Bigl({iR \over 2\pi}\Bigr)^2 + 2 \,\Tr \Bigl({iR \over 2\pi}\Bigr)^3 \Biggr] \label{fqhe6}\\
c_4 (T_cK) &=& { 1 \over 4!} \Biggl[ \Bigl(\Tr {iR \over 2\pi}\Bigr)^4 - 6 \Bigl(\Tr {iR \over 2\pi}\Bigr)^2 \,\Tr \Bigl({iR \over 2\pi}\Bigr)^2 + 8\, \Tr {iR \over 2\pi}\, \,\Tr \Bigl({iR \over 2\pi}\Bigr)^3\nonumber\\
&&\hskip .4in  +3\, \Tr \Bigl({iR \over 2\pi}\Bigr)^2 
\,\Tr \Bigl({iR \over 2\pi}\Bigr)^2 -6\,\Tr \Bigl({iR \over 2\pi}\Bigr)^4 \Biggr] \nonumber
\eeqar
Since the curvatures $R$ take values in the Lie algebra of $U(k)$, 
the traces in the above formulae are over this algebra.
Explicitly, we can write
$i R = d\omega^0 \mathbb{1} + R^a t_a$, where $\mathbb{1}$ and
$t_a$ form a hermitian basis for the $U(k)$ algebra and $t_a$ are normalized so that $\Tr (t^a t^b) = \half \delta^{ab}$. $\omega^0$ is the abelian part of the $U(k)$ spin connection.

The Chern character ${\rm ch}(V) $ in the index formula (\ref{fqhe3})
involves the gauge fields with the trace defined over the
representations of the gauge group to which the matter fields belong.
It is given by
\beq
{\rm ch}(V) = \Tr \left( e^{i {\cal F} /2 \pi} \right) = {\rm dim}\,V + \Tr ~{{i {\cal F}} \over {2\pi}} + { 1 \over 2!} \Tr~ {{i{\cal F} \wedge i {\cal F}} \over {(2\pi)^2}} + \cdots
\label{fqhe7}
\eeq 
where ${\rm dim} V$ is the dimension of the bundle $V$, i.e., the number
of matter fields.
${\cal F}$ is the gauge field strength, it includes the external
magnetic field and the $b$-fields in the present case.
(We will also be considering only Abelian gauge fields in this paper, for
simplicity, although the methodology is applicable to nonabelian fields as well.)

Since the Dolbeault index gives the degeneracy of the lowest Landau level
(LLL),
it is also the charge of the $\nu =1 $ state if we assign 
unit charge to the particles. Taking the charge density
of the LLL as $J_0$, we may then identify
\beq
{\delta S_{\rm eff} \over \delta A_0} = J_0 =  {\rm Index ~density}
\label{fqhe8}
\eeq
where the ``index density" in the above expression is the integrand of (\ref{fqhe3}).
We can then ``integrate up" from this formula to identify
most of the terms in $S_{\rm eff}$. This was the strategy used
in \cite{KN5}. The leading term in $S_{\rm eff}$ will be
a Chern-Simons term $CS(A)$ whose variational derivative with  respect to
$A_0$ is the index density. 
There can be subleading terms which correspond to
dipole and higher multipole terms in $J_0$,
which can lead to
terms involving derivatives of the fields
in the effective action; such terms are nontopological in nature. We will focus here on the topological terms. 
We have also discussed in \cite{KN5} how purely gravitational terms
can be added to this procedure. The end result is the following.

We start from the term in the Dolbeault index density in
(\ref{fqhe3}) corresponding to the $(2k+2)$-form, say,
$\I_{2k +2}$.
We can then obtain an associated Chern-Simons form
$({\Lambda})_{2k+1}$ by writing
\beq
\I_{2k +2} = {1\over 2\pi} d ({\Lambda})_{2k+1}
\label{fqhe9}
\eeq
The effective action is then given by the
integral of $({\Lambda})_{2k+1}$ over the manifold
$M \times \mathbb{R}$, where $M$ is the $2k$-dimensional spatial manifold and $\mathbb{R}$ denotes the time direction.
Written out, this is of the form
\beq
S_{\rm eff} = \int \Lambda_{2k +1}
= \int \Bigl[ {\rm td}(T_c M) \wedge \sum_p  (CS)_{2 p+1} ( A)\Bigr]_{2 k+1}
+ 2 \pi\int \Omega^{\rm grav}_{2k+1}
\label{fqhe10}
\eeq
The integrand on the right hand side of this equation identifies the explicit
form of $\Lambda_{2k+1}$.
Here $(CS)_{2 p +1}(A)$ is the Chern-Simons term associated with
just the gauge part and is defined by
\beq
{1\over 2 \pi} d (CS)_{2 p +1} = {1 \over (p+1)!} \Tr \left(
{ i \F \over 2 \pi}\right)^{p +1}
\label{fqhe11}
\eeq
One should expand the terms in the square brackets in
(\ref{fqhe10}) in powers of
curvatures and $F$ and pick out the term corresponding to
the $(2k+1)$-form. The subscript $2k+1$ for
the square brackets is meant to signify this. The purely gravitational term 
$\Omega^{\rm grav}_{2k+1}$  in (\ref{fqhe10}) is defined by
\beq
\left[ {\rm td}(T_cM) \right]_{2 k +2}
=  d\, \Omega_{2 k+1}^{\rm grav} 
\label{fqhe12}
\eeq
This method of constructing the effective action can be extended to
include 
higher Landau levels for some
special cases. It is equivalent to considering fields of nonzero spin in the lowest Landau level. However, since we do not need it for this article, 
we refer the interested reader to \cite{KN5} for details.

Turning to the spectator fields, note that 
anomalies in $2k$ dimensions are obtained from the index density
(or index polynomial) in $(2k+2)$ dimensions via the descent 
procedure, (See \cite{anom} for a general discussion of the
descent procedure.)
For the spectator fields, which are chiral spinors, we will need the Dirac index.
This is given by
\beq
{\rm Dirac ~Index}=  \int {\hat A} (M) \wedge {\rm ch} (V)
= \int {\hat A} (M) \wedge \Tr \left( e^{ i \F \over 2\pi}\right)
\label{fqhe17}
\eeq
Here ${\hat A}(M)$ is the ${\hat A}$-genus which has the expansion
\beq
{\hat A} (M) = 1 - {1\over 24} (c_1^2 - 2 c_2) + \cdots
\label{fqhe18}
\eeq
where the ellipsis denotes higher forms.
Also the gauge field will be just made of the $b$-fields,
so that $i \F = Q^{(n)}d b^{(n)}$, $Q^{(n)}$ being the charge for
coupling to $b^{(n)}$ for the spinor of interest.
\section{The parton construction in 2+1 dimensions and anomaly cancellation}
We now turn to the construction of the Laughlin states of filling fraction $\nu = 1/m$ in 2+1 dimensions
using the parton picture. We consider $m$ parton fields $q_i$, $i = 1, 2, \cdots, m$. 
The partons are coupled to the external electromagnetic field with
charge $1/m$; we absorb $e$ into the gauge field $A_\mu$. In addition, the partons couple to a set of $b$-fields. We will need at least
$(m-1)$ $U(1)$ gauge fields to ensure the equality of all the parton occupation numbers.
It is convenient to choose the charges for their coupling to the partons 
as proportional to the set of
diagonal matrices $h^{(n)}$, $n = 1, 2, \cdots, (m-1)$, in the fundamental $m\times m$ matrix representation of $SU(m)$. In particular the $h^{(n)}$ matrices can be written as
\beq
h^{(n)}= {1\over \sqrt{2 n (n+1)}} \,{\rm diag}\{ \underbrace{1, 1, \cdots, 1}_{n}, -n,0,\cdots,0 \},~~~~~n=1,2,\cdots,(m-1)
\label{matrix1}
\eeq
The matrices $h^{(n)}$ are traceless and are normalized
as $\Tr (h^{(n)} h^{(n')} ) = {\half} \delta^{n n'}$.  
Although we use this language, we do not have an $SU(m)$ gauge theory, we are only using the diagonal matrices, i.e., the Cartan subalgebra, so the gauge group for the $b$-fields is $U(1)^{m-1}$. (A generalization to using an $SU(m)$ gauge theory is possible, but not particularly useful in this context.) 

Turning to the effective action, note that in 2+1 dimensions, we need the term $\I_4$ corresponding to the 4-form in the Dolbeault index density
from (\ref{fqhe3}) with 
\beq
i \F = {d A\over m} {\mathbb{1}}  + \sum_{n=1}^{m-1} db^{(n)} h^{(n)}.
\label{fqhe13}
\eeq
 It is straightforward to see that $\I_4$ is given by
\beqar
\I_4 &=& { 1 \over 2!} \Tr~ {{i{\cal F} \wedge i {\cal F}} \over {(2\pi)^2}} ~+~ {c_1 \over 2} ~\Tr~ {i{\cal F}  \over {(2\pi)}}~+~{{c_1^2 + c_2} \over 12} ~{\rm dim} V \nonumber \\
 &=& {1 \over 2\pi} \left[ 
{1\over 4 \pi m} dA\,dA + {1 \over 4\pi} dA \, d\omega + {m \over 24 \pi} d \omega\, d \omega
+ {1\over 8\pi} \sum_{n = 1}^{m-1}db^{(n)} db^{(n)} \right]
\label{fqhe14}
\eeqar
We have used $c_1$, $c_2$ from (\ref{fqhe6}) with the simplifications
appropriate to 2+1 dimensions, i.e., $c_2 =0$, $c_1 = d\omega^0/2\pi 
\equiv d\omega /2\pi$.
The contribution to the bulk effective action from the partons is thus
\beq
S_{\rm eff}(q) = \int~{1\over 4\pi} \left[ {A dA \over m} + A d\omega 
\right] + {m \over 24 \pi} \omega d\omega + {1\over 8\pi} \sum_{n = 1}^{m-1} b^{(n)} db^{(n)}
\label{fqhe15}
\eeq
The last term is what gives rise to the anomaly for the $b$-fields on the boundary of the droplet.

\subsection{A solution for anomaly cancellation}

We now turn to the cancellation of the $b$-anomaly. 
In (2+1) dimensions the direct integration of the Chern-Simons action for the $b$-fields leads to the so-called framing anomaly \cite{framing}. 
Effectively, this introduces an additional gravitational Chern-Simons term.  
As mentioned in the Introduction, we shall follow a different strategy.
Integration over the $b$-fields is straightforward, as in any gauge theory without gauge anomalies, if we cancel out the anomaly for the $b$-fields.
This can be done
by introducing a set of auxiliary fields coupled to $b$-fields but not to
the electromagnetic fields. This method of anomaly cancellation
has the advantage of being generalizable to higher dimensions where
integrating out higher Chern-Simons forms for the
$b$-fields in a nontrivial gravitational background is not straightforward.

We will first give
a particular solution and then discuss in what sense this would be the
minimal solution. Consider $(m-1)$ spectator spinors denoted by
$\chi_i$, $i= 1, 2, \cdots, (m-1)$, which are of right-handed chirality
(i.e., of opposite chirality compared to the partons) and couple only to the $b$-fields. They have the same charges as the partons for the $b^{(n)}$ gauge fields, $n = 1, 2, \cdots, (m-2)$ and a charge 
$Q/\sqrt{2m(m-1)}$ for $b^{(m-1)}$.
The $\chi$-fields do not couple to the electromagnetic field. Therefore the corresponding $\F$ in (\ref{fqhe17}) is of the form 
\beq
i \F = \sum_{n=1}^{m-2} db^{(n)} \tilde{h}^{(n)} ~+~{Q ~db^{(m-1)} \over \sqrt{2m(m-1)}}~ {\mathbb{1}}, 
\label{fqhe18a}
\eeq
where $\tilde{h}$ are traceless $(m-1) \times (m-1)$ matrices, such that $\Tr (\tilde{h}^{(n)} \tilde{h}^{(n')} ) = {\half} \delta^{n n'}$. They are as in
(\ref{matrix1}) with $n$ ranging from $1$ to $(m-2)$. 
 (In other words they are identical to $h^{(n)}$ with the range of $n$ truncated at $(m-2)$.) Using this, we find that the four-dimensional Dirac index density for this case is
\beq
{\rm Dirac}~\I_4 =  {1 \over {2 (2\pi)^2}} \left[ \sum_{n=1}^{m-2} {1 \over 2} d b^{(n)} d b^{(n)}
+ {Q^2  \over {2 m}} d b^{(m-1)} db^{(m-1)} \right] - {(m-1) \over 24} {d\omega d\omega \over {(2\pi)^2}}
\label{4ddi}
\eeq
and the corresponding effective action for the $\chi$ fields is
\beq
S_{\rm eff} (\chi) = - \int \left[ {1\over 8\pi} \sum_{n=1}^{m-2} b^{(n)} d b^{(n)}
+ {Q^2  \over {8 \pi m}} b^{(m-1)} db^{(m-1)}
\right] + (m-1 )\int {\omega d\omega\over 48\pi}
\label{fqhe19}
\eeq
The last term in this expression is the purely gravitational contribution from
${\hat A} (M)$.

Notice that the anomalous terms for $b^{(n)}$ for
$n = 1, 2, \cdots, (m-2)$, cancels out between 
(\ref{fqhe15}) and (\ref{fqhe19}). In order to cancel the anomalous term for $b^{(m-1)}$ we introduce two additional spinors,
one of left chirality denoted by $\chi'$ and one of right chirality
denoted by $\chi''$.
These fields couple to $b^{(m-1)}$ and gravity.
Their $b^{(m-1)}$ charges will be $\alpha/\sqrt{2 m (m-1)}$
for $\chi'$ and $\beta/\sqrt{2 m (m-1)}$ for $\chi''$.
Since we have one of each chirality, the gravitational contribution from
${\hat A}(M)$ will cancel out and we find
\beq
S_{\rm eff} (\chi', \chi'') =  \int {\alpha^2 -\beta^2\over {2 m (m-1)}} \,{1\over 4\pi} b^{(m-1)} db^{(m-1)} 
\label{fqhe20}
\eeq
 Combining terms (\ref{fqhe15}), (\ref{fqhe19}) and (\ref{fqhe20}) we find
the anomaly cancellation condition
\beq
{1\over 8\pi} b^{(m-1)} d b^{(m-1)} \left[
1- {Q^2 \over m} + {\alpha^2 \over m (m-1)} - {\beta^2 \over m (m-1)}
\right] = 0
\label{fqhe21}
\eeq
This has the solution
\beq
Q = \pm (2 l+1), \hskip .2in \alpha = \pm 2l (l+1) , \hskip .2in
\beta = \pm 2 l^2 , \hskip .2in m = 2 l +1
\label{fqhe22}
\eeq
where we write $m = 2 l +1$, since it is an odd integer.

The choice of a particular sign for $\alpha$, $\beta$ in
this solution can be motivated by consideration of possible composite fields.
The electron or the physical fermion is made up of the partons, so
it can be represented by the composite field
\beq
\psi  \sim q_1 q_2 \cdots q_m
\label{fqhe23}
\eeq
This field $\psi$
has the correct electric charge, and zero charge for all
the $b$-fields, i.e., for $b^{(n)}$, $n = 1, 2, \cdots, (m-1)$. 
The $b$-fields could bind the spectator fields as well; for example, one
could make a composite from all the $\chi$'s of the form
\beq
\Xi  \sim \C_{r_1 r_2 \cdots r_{m-1}}\, {\bar\chi}^{r_1} {\bar\chi}^{r_2}
\cdots {\bar\chi}^{r_{m-1}} \, \chi'_s \, {\bar\chi}''^s
\label{fqhe24}
\eeq
for a suitable choice of coefficients $\C_{r_1 r_2 \cdots r_{m-1}}$.
This field has zero charge for all $b^{(n)}$, $n = 1, 2, \cdots, (m-2)$.
The charge carried by
$\Xi$ for
the $b^{(m-1)}$-field is
\beq
Q^\Xi_{m-1} = {{\alpha - \beta - Q (m-1)}\over \sqrt{2 m (m-1)}}
\label{fqhe25}
\eeq
If we require this to be zero as well,
a consistent choice of signs in the solution
(\ref{fqhe22}) is
\beq
Q = 2l +1, \hskip .2in \alpha = 2l (l+1) , \hskip .2in \beta = - 2 l^2
\label{fqhe26}
\eeq
The spectator fields are thus
$\chi_i$, $\chi'$, $\chi''$ with the charges for coupling to $b^{(m-1)}$ as
given in this equation.

We can now collect all the remaining terms in the effective action
from (\ref{fqhe15}), (\ref{fqhe19}) and (\ref{fqhe20}) as
\beqar
S_{\rm eff} &=& \int {1\over 4\pi} \left[ {A dA \over m} + A d\omega 
 + \left({m \over 4 }-{1 \over 12 } \right) \omega d\omega \right] \nonumber\\
&=& \int {1 \over 4\pi m} \left( A  + {\half} m \omega\right)
d  \left( {A } +  {\half} m \omega \right)
- {1 \over 48 \pi}\int \omega d\omega
\label{fqhe27}
\eeqar
This result agrees with what has been calculated for the $\nu= 1/m$
Laughlin states \cite{{framing},{AG2}}. Notice that the Hall conductivity
(defined by the functional derivative of $S_{\rm eff}$ with respect to
$A_i$) has the expected $\nu = 1/m$ behavior. The Wen-Zee 
term is the same as for the $\nu =1 $ case.
The gravitational term is often split as
shown in the second line of (\ref{fqhe27}),
with the last term of the second line referred to as the
``pure gravitational anomaly" or even just the ``gravitational anomaly".
Our calculation shows the precise mathematical sense in which this
distinction is to be made.
As shown in \cite{gilkey}, the Dolbeault index can be written as
\beq
{\rm Index}( \bD ) = \int_M {\rm td}(T_cM) \wedge {\rm ch} (V)
= \int_M  {\hat A}(M) \wedge \Tr \left( e^{{i \F \over 2\pi}+ {i R \over 4 \pi}}
\right)
\label{fqhe28}
\eeq
Since 
\beq
{i \F \over 2 \pi} + {i R \over 4 \pi}
= {1\over m} d \bigl(A + {\half} m \omega \bigr),
\label{fqhe28a}
\eeq
the Chern character shows how the combination
of the gauge fields and the spin connection as
$(A + {\half} m\omega)$
 arises in a natural way.
For the partons we get $m$ times the contribution from the
${\hat A}$-genus, while the $\chi$-fields give $-(m-1)$ times the same
contribution, leading to the last term in the second line
of (\ref{fqhe27}).
Thus the ``pure gravitational anomaly" can be understood as the
contribution of the ${\hat A}$-genus.

An alternative solution for anomaly cancellation would be to take
$m$ Dirac spinors for the spectators with $b$-charges which are identical to the charges of the $m$ partons.
This would obviously cancel the $b$-anomalies but would give a
different value for the purely gravitational term in the final effective action.
In particular, the action will not agree with the action for the integer QHE
when we set $m = 1$. Recall that the action for the integer case
does not require spectators (or partons) and is unambiguously determined
by the Dolbeault index density. Requiring that the anomaly cancellation 
should be consistent with that result (upon setting $m = 1$)
rules out this alternative cancellation solution.

\subsection{Transport coefficients}
We can now read off the transport coefficients, the Hall conductivity and the Hall viscosity, from the effective action (\ref{fqhe27}). 
The variations of the effective action with respect to the
electromagnetic field $A$ and the metric $g_{ml}$ give the
current $J^i$ and the energy-momentum tensor
$T^{ml}$ as
\beq
\delta S_{\rm eff} =  \int d^{2k+1} x\, \sqrt{\det g}\,
\left[ J^i \delta A_i - {1\over 2} T^{ml} \delta g_{ml} \right]
\label{transport}
\eeq
In the present case, the Hall current is given by
\beq
J^i = \epsilon^{ij} \left[ {E_j \over 2\pi m} + {R_{j0} \over 4 \pi} \right]
\label{Hcurrent}
\eeq
where $E_i = F_{i0}$. The Hall conductivity is $\sigma_H = {\nu \over 2\pi}$. An interesting feature of (\ref{Hcurrent}), which is also valid in all higher dimensions \cite{KN6}, is that a Hall current can be generated from 
time-variation of the metric even if there is no external electric field applied to the system. 

In order to identify the Hall viscosity we have to derive the energy-momentum tensor from the effective action (\ref{fqhe27}) and further identify the term involving the time-derivative of the metric. A detailed calculation for the Hall viscosity in two and four dimensions for $\nu =1$ was done in  \cite{KN6}. It is straightforward to modify those results for $m \ne 1$. We then find
\begin{align}
T^{m l} =   {1 \over {8 \pi\sqrt{\det g}}} \left( g^{mi} \e^{lk}+ g^{li} \e^{mk}  
 \right)
\biggl\{&\left[ { B \over 2} + \left({m \over 4} -{1 \over 12}\right) \left( {R \over 2} -\nabla^2 \right)\right]  {\dot g}_{ki}
 \nonumber \\
&+ {1 \over 2} \left( {m \over 4} -{1 \over 12}\right) \nabla_i \nabla_k \left(g^{rn} {\dot g}_{rn} \right) \biggr\}
\label{119}
\end{align}
where $R$ is the Ricci scalar curvature and the
magnetic field $B$ given by
\beq
F_{ij} = \epsilon_{ij} ~B~\sqrt{\det g}
\label{120}
\eeq
Comparing (\ref{119}) with the expression of the energy-momentum tensor in terms of the Hall viscosity, we see that we can write 
\beqar
\sqrt{\det g}~T^{m l} &=&   {1\over 2} \, \eta_{H} \, \left( g^{mi} \e^{lk}+ g^{li} \e^{mk}  
 \right)  {\dot g}_{ki}
 \nonumber \\
&&+ {1 \over 2}\, \eta^{(2)}_{H} \,\left( g^{mi} \e^{lk}+ g^{li} \e^{mk}  
 \right) \nabla_i \nabla_k \left(g^{rn} {\dot g}_{rn} \right) 
\label{120}
\eeqar
where the coefficients $\eta_H$ and $\eta_H^{(2)}$ can be read off as
\beqar
\eta_H &=& {1 \over 4\pi} \left[ { B \over 2} + \left( {m \over 4} -{1 \over 12}\right) \left( {R \over 2} + \vec{k}^2 \right) \right]\nonumber \\
\eta^{(2)}_{H} & =& { 1 \over 8\pi} \left( {m \over 4} -{1 \over 12}\right)
\label{121}
\eeqar

We will close this subsection with a few clarifying remarks.
Although we have focused on the effective action and
anomaly cancellation, it is important to consider the occupation numbers
for the partons. In the case of a two-dimensional spatial manifold, this will be given by the integral of
the 2-form from the index polynomial (which will also be
the variation of the action with respect to $A_0$).

The occupation number for the fully filled LLL for each of the partons then takes the form
\beq
n_i = \int \left[ {F \over 2\pi m} + {R \over 4 \pi}\right]  + \int
(h^{(n)})_i {d b^{(n)}  \over 2\pi}
\label{fqhe30}
\eeq
where we have separated out the part depending on the $b$-fields.
Notice that this means that $b^{(1)}$ couples to
$q_1$ and $q_2$, with charges $\pm {\half}$, but not to the other
partons.
Similarly, $b^{(2)}$ couples to $q_1$, $q_2$ and $q_3$
(with charges ${1\over\sqrt{12}}$, ${1\over\sqrt{12}}$, $-{2\over\sqrt{12}}$), but 
not to others, etc.
For equality of the $n_i$, we therefore need to require that
\beq
\int db^{(n)} = 0
\label{fqhe31}
\eeq
for all $n$. This does not in any way come into the question of the anomaly cancellation since (\ref{fqhe31}) does not imply
$\int b^{(n)} d b^{(n)} = 0$.

Our second remark relates to the nature of the composite
field $\Xi$ in (\ref{fqhe24}).
While the spectator fields are introduced for anomaly cancellation,
composites like $\Xi$ have zero electric charge. 
So there is no measurable response for them from changes in
external electromagnetic fields. They do respond to gravitational
perturbations, but this is already captured by the effective action.
Thus, beyond effects derivable from
the effective action in (\ref{fqhe27}), composites
of the spectator
fields such as $\Xi$ are more or less irrelevant.

\subsection{A concern for other values of $\nu$}

Finally, we point out a difficulty with the parton picture for states with filling fractions different from $\nu = 1/m$ when one deals with curved manifolds. 
This can be illustrated by a simple example, say, for $\nu = 2/5$. 
The usual parton picture for this uses three partons, with charge $2/5$ for $q_1$, $q_2$ and charge $1/5$ for
$q_3$. The partons $q_1$, $q_2$ fill the LLL, $\nu=1$, while $q_3$ fills the LLL and
the first excited level, $\nu=2$.
With the $b$-charge assignments as in (\ref{matrix1}), (\ref{fqhe13}), we find \cite{KN5, AG1}
\beqar
S_{\rm eff}(q)&=&{1\over 4\pi} \Bigg\{
\left( {2\over 5} A + {b^{(1)}\over 2} + {b^{(2)}\over \sqrt{12}}
+ {\omega \over 2} \right) \,d \left( {2\over 5} A + {b^{(1)}\over 2} + {b^{(2)}\over \sqrt{12}}
+ {\omega \over 2} \right) - {1\over 12} \omega d\omega\nonumber\\
&&\hskip .25in + \left( {2\over 5} A - {b^{(1)}\over 2} + {b^{(2)}\over \sqrt{12}}
+ {\omega \over 2} \right) \,d \left( {2\over 5} A - {b^{(1)}\over 2} + {b^{(2)}\over \sqrt{12}}
+ {\omega \over 2} \right) - {1\over 12} \omega d\omega\nonumber\\
&&\hskip .25in +\left( {1\over 5} A - 2 {b^{(2)}\over \sqrt{12}}
+ {\omega \over 2} \right) \,d \left( {1\over 5} A - 2 {b^{(2)}\over \sqrt{12}}
+ {\omega \over 2} \right) - {1\over 12} \omega d\omega\nonumber\\
&&\hskip .25in +\left( {1\over 5} A - 2 {b^{(2)}\over \sqrt{12}}
+ {3\omega \over 2} \right) \,d \left( {1\over 5} A - 2 {b^{(2)}\over \sqrt{12}}
+ {3\omega \over 2} \right) - {1\over 12} \omega d\omega\Bigg\}\nonumber\\
&=&{1\over 4\pi} \Bigg\{ {2\over 5} A dA + {8 \over 5} A d \omega + 
{8\over 3} \omega d\omega - {6 \over \sqrt{12}} b^{(2)} d\omega
+ {1\over 2} b^{(1)} d b^{(1)} + {5 \over 6} b^{(2)} d b^{(2)}\Bigg\}
\label{fqhe32}
\eeqar
The $A dA$ and $A d \omega$ terms agree with the corresponding terms in the effective action for $\nu=2/5$ in \cite{framing}. 
There is a discrepancy for the $\omega d \omega$ term, but there will be
similar terms from the $\chi$-fields after the anomaly cancellation, so this is not yet important.
However a problem with this choice of partons is already apparent at this stage, before we even consider anomaly cancellation.
For the number of partons of each kind,
we find
\beqar
n_1&=& {1\over 2\pi} \int 
\left[ {2\over 5} d A + {1\over 2} d \omega + {d b^{(1)} \over 2} 
+ {d b^{(2)} \over \sqrt{12}} \right]\nonumber\\
n_2&=& {1\over 2\pi} \int 
\left[ {2\over 5} d A + {1\over 2} d \omega - {d b^{(1)} \over 2} 
+ {d b^{(2)} \over \sqrt{12}} \right]\nonumber\\
n_3&=& {1\over 2\pi} \int 
\left[ {2\over 5} d A + {4\over 2} d \omega 
-4  {d b^{(2)} \over \sqrt{12}} \right]
\label{fqhe33}
\eeqar
These are given by varying the action (\ref{fqhe32}) with respect to
$(2/5) A_0$ for $q_1$, $q_2$ and with respect to
$(1/5)A_0$ for $q_3$. Alternatively, these can be obtained directly from the
two-dimensional index theorem.

As should be clear from (\ref{fqhe33}), for
curved spaces, such as the sphere for example, the occupation numbers for the partons will not be equal to each other, which will be problematic in writing down the many-body electron wave function in terms of partons.
This is true for all cases where $\int d\omega \neq 0$, although
we can still use the action (\ref{fqhe32}) for manifolds with small perturbations of the metric around a background with $\int d\omega = 0$.
Another possibility is to choose nontrivial backgrounds with $d b \neq 0$
so as to match the numbers in (\ref{fqhe33}). We do not pursue these issues in detail here, our aim is to point out that there are subtleties in extending
a parton picture to manifolds of nontrivial geometry and topology.
The case of $\nu = 1/m$ discussed above is not affected by this issue.

\section{The parton construction in 4+1 dimensions}
\subsection{Anomaly cancellation and the effective action}
The construction for the fractional states in 4+1 dimensions is
fairly straightforward since the method is clear from the
$(2+1)$-dimensional case worked out in detail in the last section. 
Towards this, we again consider $m$ partons, each of charge $1/m$,
and coupled to $b$-fields with charges as given in 
(\ref{matrix1}). Each type of parton will be in a $\nu =1$ state.

The 6-form Dolbeault index density can be identified from (\ref{fqhe3}), where 
$i \F = (d A/m)~ {\mathbb{1}}+ \sum_{n=1}^{m-1} db^{(n)} h^{(n)}$. 
\beq
\I_6 = { 1 \over 3!} \Tr~ {{i{\cal F} \wedge i {\cal F} \wedge i {\cal F}} \over {(2\pi)^3}} ~+~ {c_1 \over 4} ~\Tr~ {{i{\cal F} \wedge i {\cal F}}  \over {(2\pi)^2}}~+~{{c_1^2 + c_2} \over 12} ~\Tr~{i{\cal F} \over 2\pi}~ +~ {c_1 c_2 \over 24} {\rm dim} V 
\label{sixind}
\eeq
The corresponding effective action for the partons is then
\beqar
S_{\rm eff} (q) &=&{1\over 24 \pi^2 m^2 }  A F^2 
+ {c_1 \over 8\pi m} A F  + {c_1^2 + c_2 \over 12} A
+ {m \over 192\pi^2} \Tr \omega \left[ (\Tr d\omega)^2 - \Tr (R R) \right]
\nonumber\\
&& +\left[ {1\over 16 \pi^2 m} F + {c_1 \over 16\pi} \right]
\sum_{n=1}^{m-1} b^{(n)} db^{(n)}\nonumber\\
&& + {1\over 24\pi^2} \sum_{n, n', n''=1}^{m-1} b^{(n)} db^{(n')} db^{(n'')} \Tr \left(h^{(n)} h^{(n')} h^{(n'')} \right)
\label{4d-2}
\eeqar
For ease of working out the anomaly cancellation, it is useful to
separate out the term involving the $b^{(m-1)}$-field in the last  term
of this expression. We find
 \begin{align}
 &{1\over 24\pi^2}\sum_{n, n', n''=1}^{m-1} b^{(n)} db^{(n')} db^{(n'')} \Tr \left(h^{(n)} h^{(n')} h^{(n'')} \right)\nonumber\\
&= {1\over 24\pi^2}\Bigg\{\sum_{n,n',n''}^{m-2} b^{(n)} db^{(n')} db^{(n'')} \Tr \left(h^{(n)} h^{(n')} h^{(n'')} \right) \nonumber\\
&+ {3 \over 2} {b^{(m-1)}\over \sqrt{2 (m^2 - m)}}\sum_{n=1}^{m-2} b^{(n)} db^{(n)} 
- m (m-1) (m-2) {b^{(m-1)} d b^{(m-1)} d b^{(m-1)} \over 
 (2 (m^2 -m))^{3\over 2}}\Bigg\}
 \label{4d-3}
 \end{align}

Turning to the cancellation of anomalies, we consider
$(m-1)$ right-chiral spinors $\chi$ with $b$-charges as in (\ref{fqhe18a}) for
$n = 1, 2, \cdots, (m-2)$. As in the two-dimensional case the $\chi$ spinors carry a charge of $Q /\sqrt{2 (m^2 -m)}$ for coupling
to $b^{(m-1)}$
and they do not couple to electromagnetism (i.e., have zero electromagnetic charge).

The relevant Dirac index density in 6-dimensions is
\beq
{\rm Dirac~ index} =  { 1 \over 3!} \Tr~ {{i{\cal F} \wedge i {\cal F} \wedge i {\cal F}} \over {(2\pi)^3}} ~-~{{c_1^2 -2 c_2} \over 24} ~\Tr~{i{\cal F} \over 2\pi}
\label{dsixind} 
\eeq
where $i \F = \sum_{n=1}^{m-2} db^{(n)} \tilde{h}^{(n)} ~+~{Q \over \sqrt{2m(m-1)}} db^{(m-1)}~ {\mathbb{1}}$ and $\tilde{h}$ are traceless $(m-1) \times (m-1)$ matrices, such that $\Tr (\tilde{h}^{(n)} \tilde{h}^{(n')} ) = {\half} \delta^{n n'}$, as in (\ref{fqhe18a}).

The corresponding effective action for $\chi$ is
\beqar
S(\chi ) &=&- {1\over 24 \pi^2} \Bigg\{\sum_{n,n',n''}^{m-2} b^{(n)} db^{(n')} db^{(n'')} \Tr \left(h^{(n)} h^{(n')} h^{(n'')} \right) \nonumber\\
 &&+ {3 Q \over 2} {b^{(m-1)}\over \sqrt{2 (m^2 - m)}} \sum_{n=1}^{m-2} db^{(n)}~db^{(n)}
+  (m-1) Q^3\, {b^{(m-1)} d b^{(m-1)} d b^{(m-1)} \over 
 (2 (m^2 -m))^{3\over 2}}\Bigg\}\nonumber\\
 && + Q \, {m-1\over 24} (c_1^2 - 2 c_2)  
{b^{(m-1)} \over 
 \sqrt{2 (m^2 -m)}}
 \label{4d-4}
 \eeqar
We will also introduce $N$ left-chiral spinors, denoted by $\chi'$, coupled to
$b^{(m-1)}$ with charge ${\alpha \over \sqrt{2m(m-1)}}$ and $M$ right-chiral spinors
$\chi''$
coupled to $b^{(m-1)}$ with charge ${\beta \over \sqrt{2m(m-1)}}$.
The contribution of these will be
\beqar
S(\chi') &=& {1\over 24 \pi^2} N\alpha^3 \,{b^{(m-1)} d b^{(m-1)} d b^{(m-1)} \over 
 (2 (m^2 -m))^{3\over 2}} - {1\over 24} (c_1^2 - 2 c_2 ) N \alpha
 {b^{(m-1)}\over \sqrt{2 (m^2 - m)}}\nonumber\\
 S(\chi'') &=& -{1\over 24 \pi^2} M\beta^3 \,{b^{(m-1)} d b^{(m-1)} d b^{(m-1)} \over 
 (2 (m^2 -m))^{3\over 2}} + {1\over 24} (c_1^2 - 2 c_2 ) M \beta
 {b^{(m-1)}\over \sqrt{2 (m^2 - m)}}
 \label{4d-5}
 \eeqar
 Combining (\ref{4d-2})-(\ref{4d-5}), we see that the leading anomaly, the cubic b-term, cancels if  
 \beqar
 && Q-1 = 0 \nonumber\\
 && -m(m-1)(m-2) + N \alpha^3 - (m-1) Q^3 - M \beta^3 =0
  \label{4d-6}
 \eeqar
 Further the linear b-term, as well as the total $b^{(m-1)}$ charge, cancels if 
 \beq
 N\alpha - (m-1) Q - M\beta = 0
 \label{4d-charge}
 \eeq
We will consider two possible solutions
to (\ref{4d-6}), (\ref{4d-charge}), given by
\beqar
{\rm I}.~~&&Q =1, \hskip .2in N =0, \hskip .2in M =1, \hskip .2in
\beta = - (m-1) \label{4d-7}\\
{\rm II}. ~~&&Q =1, \hskip .2in N =1, \hskip .2in M =1, \hskip .2in \alpha =0, \hskip .2in
\beta = - (m-1)
\label{4d-7a}
\eeqar
For the first solution, the $m$ spectator fields 
$\chi$, $\chi''$ have $b$-charges which are identical to the
$b$-charges for the partons (and there is no $\chi'$ field).
The second solution is essentially what we had in 2+1 dimensions,
with one $\chi'$ and one $\chi''$ field in addition to $(m-1)$
$\chi$'s.
Notice that, as far as $b$-anomalies are concerned, these two
solutions have the same content, since $\alpha =0$ for 
the second solution.
In principle, we could also add an equal number of left and right chirality
fields whose anomalies will cancel against each other, but
(\ref{4d-7}), (\ref{4d-7a}) represent the minimal choice.

As in the case of two dimensions, the 
electron will correspond to a composite field
\beq
\psi   \sim q_{1} q_{2} \cdots q_m
\label{4d-8}
\eeq
There can also be composites made of the
spectator fields, but, for reasons given after
(\ref{fqhe31}), these are not important for us.

Adding the action for the partons (\ref{4d-2}) and the action for the right chiral spinors $\chi, \chi''$ (\ref{4d-4}), (\ref{4d-5})  we find an effective action where the leading anomalous term, the cubic-$b$ term, is cancelled. 
(Notice that $\chi'$ with $\alpha = 0$ does not contribute to the
action.) The resulting effective action takes the form
\beqar
S_{\rm eff} &=& {1\over 24 \pi^2 m^2 }  A F^2 
+ {c_1 \over 8\pi m} A F  + {c_1^2 + c_2 \over 12} A
+ {m \over 192\pi^2} \Tr \omega \left[ (\Tr d\omega)^2 - \Tr (R R) \right]
\nonumber\\
&& + \left[ {1\over 16 \pi^2 m} F + {c_1 \over 16\pi} \right]
\sum_{n=1}^{m-1}b^{(n)}db^{(n)}
\label{4d-9}
\eeqar
The expression above is the bulk effective action for fractional Hall states in $4+1$ dimensions.
The expression still involves the $b$ fields. Unlike in $2+1$ dimensions, we cannot resort to known framing-anomaly-type results to integrate them out. However, we can still eliminate them as follows.

The final term in the effective action can be rewritten as
\begin{align}
\left[ {1\over 16 \pi^2 m} F + {c_1 \over 16\pi} \right]
\sum_{n=1}^{m-1}b^{(n)}db^{(n)} &= \left[ {1\over 16 \pi^2} \left( {A\over m} + {\Tr \omega \over 2}\right)
\sum db^{(n)}db^{(n)} \right] \nonumber\\
&+ d \left[{1\over 16\pi^2}  \left( {A \over m} + {\Tr \omega \over 2}\right) 
\sum b^{(n)}db^{(n)}\right]
\label{4d-10}
\end{align}
The second term on the right hand side is a total derivative and
integrates to a term on the
boundary. Since it is a local term involving the gauge fields on the boundary,
it can be removed by a choice of regularization when the spectator fields are integrated out. The bulk term is thus invariant under gauge transformations
of the $b$-fields, so there are no more obstructions to integrating out the
$b$-fields. Notice also that $A$ and $\omega$ are coupled to a current which is conserved in the bulk. In fact, the first term on the right hand side of
(\ref{4d-10}) can be written as
\beq
{1\over 16 \pi^2} \left( {A\over m} + {\Tr\, \omega \over 2}\right)
\sum db^{(n)}db^{(n)} 
= \left( {A_\mu\over m} + {\Tr\, \omega_\mu \over 2}\right) \, {\cal J}^\mu
\label{4d-10a}
\eeq
where ${\cal J}^\mu$ is the dual of ${1\over 16 \pi^2}\sum db^{(n)}db^{(n)}$. 
By construction ${\cal J}^\mu$ is a conserved current.
When the $b$-fields are integrated out, this term will produce
powers of $A$ and $\omega$ coupled to correlators of the current
${\cal J}^\mu$. In general, such correlators are not topological
but since ${\cal J}^\mu$ is conserved, they will have 
appropriate transversality properties so that the result will involve
only $dA$'s and $d \omega$'s. Further, the expectation
value of just one power of the current should be zero, indicating that
the monopole moment of the charge should be zero on average.
Thus we expect that these nontopological terms are of the dipole or 
higher multipole nature.
Setting aside these nontopological terms, the effective action
is thus reduced to
\beq
S_{\rm eff} = {1\over 24 \pi^2 m^2 }  A F^2 
+ {c_1 \over 8\pi m} A F  + {c_1^2 + c_2 \over 12} A
+ {m \over 192\pi^2} \Tr \omega \left[ (\Tr d\omega)^2 - \Tr (R R) \right]
\label{4d-11}
\eeq
Using (\ref{fqhe6}) we can rewrite (\ref{4d-11}) as
\beqar
S_{\rm eff} &=&{ 1 \over {(2\pi)^2}} \int \Biggl\{ { 1 \over 3!~m^2} \Bigl({A }+ m \omega^0\Bigr) \Bigl[d\Bigl({A}+ m \omega^0\Bigr)\Bigr]^2 \nonumber \\
&&-{ 1 \over 12} \Bigl({A }+ m \omega^0\Bigr) \Biggl[  (d\omega^0)^2 + {1 \over 4} {R^a} \wedge { R^a} \Biggr] \Biggr\}
\label{122}
\eeqar
where, for a complex manifold with holonomy
$U(k)$, we can write $R$ in terms of the $U(1)$ and $SU(k)$ components as
$R = d\omega^0 \mathbb{1} + R^a t_a$.
\subsection{Transport coefficients}
We now turn to the transport coefficients.
For $m=1$, the $(4+1)$-dimensional action
(\ref{122}) describing the bulk dynamics of the integer QHE and its  corresponding transport coefficients were studied in detail in \cite{KN6}. Here we quote those results appropriately modified for the  $m \ne 1$ case. The Hall current takes the form
\beq
J^i = {1 \over 2} {1 \over {(2\pi)^2}} \epsilon^{ijkl} {E_j \over m^2} \left( {F_{kl}} + m {\Tr\, R_{kl} \over 2} \right)
\label{4d-12}
\eeq
where we have neglected terms involving time-derivatives of the metric.
The Hall conductivity, defined by the term proportional to the electric field
$E_j = F_{j0}$, can be identified as
\beq
\sigma^{ij}_{\rm H} =  {1\over 8\pi m^2} \epsilon^{ijkl} \left( {F_{kl}} + m {\Tr\, R_{kl} \over 2} \right)
\label{4d-13}
\eeq

In order to identify the Hall viscosity one has to obtain the energy-momentum tensor. The coefficient of the term proportional to the time-derivative of the metric in the expression for $T_{i j}$ will give the Hall viscosity. Following the calculation done in \cite{KN6} and extending it to the case where $m \ne 1$, we find that the energy momentum tensor derived from (\ref{4d-11}) involves two terms,
\beq
T^{ml}  =  T_1^{ml} + T_2^{ml}
\eeq
where
\beqar
T_1^{ml} & = & {1 \over {8 (2\pi)^2}} \left(g^{mn} (J^0)^{lj} + g^{ln} (J^0)^{mj} \right)~{\dot g}_{nj} K^0 ~+~ \cdots  \label{140a} \\
d^4x\,\sqrt{\det\, g}\, K^0
&=& \left[ {1 \over 2 m} dA dA  +  {dA } d\omega^0 +
{m\over 2} d\omega^0\, d\omega^0 +{m\over 48}
R^{\alpha\beta} R^{\beta\alpha} \right]
\nonumber
\eeqar
where $\omega^0 = {1 \over 4} \epsilon^{\a\b} \omega^{\a\b}$, $R^{\a\b}R^{\b\a} = -4 R^0R^0 - R^aR^a$
and
\beq
T_2^{ml} = - { 1\over {96 (2\pi)^2}} \left(g^{mn}  (R_{rs})^{lj}+ g^{ln} (R_{rs})^{mj} \right)~{\dot g}_{nj}
\del_p \left( {A_q } + m \omega^0_q \right) { \e^{rspq} \over \sqrt{\det g}}~+~\cdots
\label{138a}
\eeq
The ellipsis in (\ref{140a}) and (\ref{138a}) refers to momentum-dependent terms of the form $\partial {\dot g}$. 
The antisymmetric tensor $(J^0)^{lj}$ in (\ref{140a}) is defined in terms of the inverse frame fields $e^{-1l\alpha}$ and the antisymmetric tensor 
$\epsilon^{\alpha\beta}$
\beq
(J^0)^{lj} = e^{-1 l \a} e^{-1j \b} \epsilon^{\a\b}
\label{106}
\eeq
The antisymmetric tensor $\epsilon^{\a\b}$ is defined so that $\epsilon^{12}=\epsilon^{34}=1$; ~$\epsilon^{13}=\epsilon^{24}=0$.

Expressions (\ref{140a}) and (\ref{138a}) give the momentum-independent terms of the Hall viscosity in 4d. As should be clear
from these expressions, the tensorial structure in the case of a curved manifold
is rather involved.
However, there is simplification for zero curvature.
In the flat limit, the 4d complex manifold decomposes into $\mathbb{C} \times \mathbb{C}$, corresponding to the
planes $(1,2)$ and $(3,4)$ where there is a constant magnetic field $B_1, B_2$ for each plane. Since the curvature terms vanish in this limit the contribution from $T_2^{ml}$ is zero. Further, we can write
$(J^0)^{lj} \rightarrow \epsilon^{ij}$ and the contribution from $T_1^{ml}$ is of the form
\beq
T^{ml} = { 1 \over {8(2\pi)^2}} \left( g^{mi} \e^{lk}+ g^{li} \e^{mk} \right) {B_1 B_2 \over m}~{\dot g}_{ki} 
\label{139}
\eeq
Comparing with (\ref{120}) we find that the Hall viscosity in this limit is
\beq
\eta_H = {1 \over 4m}  { B_1 B_2 \over (2\pi)^2}
\label{140}
\eeq
Notice that the leading term in the Hall conductivity
in (\ref{4d-13}) behaves as $1/m^2$ while the leading term of
the Hall viscosity in (\ref{140}) behaves as $1/m$.

\subsection{FQHE on $S^2 \times S^2$ and dimensional reduction}

An interesting special case to consider is the complex manifold $S^2 \times S^2$. In that case the curvature and spin connection decompose in terms of the appropriate quantities on each sphere, i.e.
\beq
\begin{pmatrix}
d\omega_1 & 0\\
0 & d\omega_2
\end{pmatrix} = d\omega^0~{\mathbb{1}} + R^a t_a
\label{matrix}
\eeq
where $d \omega^0 = \half (d \omega_1 + d \omega_2)$ and $R^3 =d \omega_1 - d \omega_2$, $R^1=R^2=0$. 

Given the above expressions we find that for $S^2 \times S^2$
\beqar
c_1 &=& { \Tr R \over 2\pi} = { d \omega_1 + d \omega_2 \over 2\pi} \nonumber\\
c_2 & =& { {(\Tr R)^2 - \Tr R \wedge R} \over {2 (2\pi)^2}} = {{ d\omega_1 d{\omega_2}} \over (2\pi)^2}
\label{s2s2}
\eeqar
It is interesting to notice that, if we assume that the electromagnetic interactions reside only on the first sphere, the (4+1)d bulk action (\ref{4d-11}) for $m=1$ (noninteracting case) dimensionally reduces to the (2+1)d bulk action (\ref{fqhe27}) for $m=1$. This is obtained by integrating over the second sphere using
\beq
\int_{S^2} ~{d\omega_2 \over 2\pi} = 2
\label{s2}
\eeq 
Similarly for $m\ne 1$, if we assume that the gauge fields such as $A$ and $b$'s reside only on the first sphere, the (4+1)d parton effective action (\ref{4d-2}) (partons are at $\nu=1$) dimensionally reduces to the (2+1)d parton effective action (\ref{fqhe15}). The dimensional reduction however does not go through at the level of the $m \ne 1$ total effective actions
(\ref{122}) and (\ref{fqhe27}) since the gravitational contribution of the chiral edge spinors is very different in (2+1)d and (4+1)d.
This is understandable since inter-particle (inter-parton) interactions are important for FQHE, in particular between the two spheres in the present case of $S^2 \times S^2$, so a naive reduction to FQHE on one of the spheres is not to be expected. 

One can further calculate the corresponding transport coefficients on $S^2 \times S^2$. The Hall conductivity is as in (\ref{4d-13}). Regarding the Hall viscosity and keeping only the momentum-independent terms, we find the following contributions from the energy momentum tensors $T_1^{12}$ and $T_2^{12}$,
\beqar
\eta_{H,1} &=& {1 \over 4(2\pi)^2} \left[ {B_1 B_2 \over m} + {1 \over 4} (B_1 R_2 + B_2 R_1) + {m \over 16} R_1 R_2 \right] \nonumber \\
\eta_{H,2} &=& -{1 \over 4(2\pi)^2} {1 \over 24} (R_1 B_2 + {m \over 2} R_1 R_2) 
\label{Hall}
\eeqar
where $R_{1,2}$ are Ricci scalars. The calculation of the Hall viscosity from the $T_1^{34}$ and $T_2^{34}$ will give similar expressions with $R_1 \longleftrightarrow R_2$.

The total contribution for the Hall viscosity from $(T_1^{12}+T_2^{12})$ is
\beq
\eta_{H} = {1 \over 4(2\pi)^2} \left[ {B_1 B_2 \over m} + {1 \over 4} (B_1 R_2 + B_2 R_1) - {1 \over 24} B_2 R_1+ {m \over 24} R_1 R_2 \right] 
\label{total Hall}
\eeq
For $m=1$, it is straightforward to check that the limit where $B_2 = 0$ and $\int R_2 = 8 \pi$ indeed produces the (2+1)d expression for the Hall viscosity for $m=1$, eq. (\ref{121}),  confirming the dimensional reduction situation for $m=1$ mentioned earlier.

\section{Concluding remarks}
It is useful to recapitulate briefly the arguments and results of this paper,
since a number of necessary but ancillary comments were made along the way and the main thread of logic may not have been easy to follow.
The basic idea is to generalize the parton picture which has been used to construct fractional quantum Hall states in two spatial dimensions.
The electron is viewed as a composite particle made of several partons
with auxiliary gauge fields (the $b$-fields) binding them together.
The partons are in quantum Hall states of integer filling fraction; this state,
viewed
in terms of the electron, is a fractional quantum Hall state.
The action for the $b$-fields involves terms of the Chern-Simons type
in 2+1 dimensions. One can integrate them out to get a
gravitational CS term, the action for the so-called framing anomaly.
While this is fairly straightforward in 2+1 dimensions,
a similar procedure in higher dimensions would lead to higher
CS forms and this leads to an impasse
since integrating out CS theories
 in higher dimensions is still not well understood.
 However, we notice that one can introduce a set of auxiliary fields
 to cancel out any gauge anomaly for the $b$-fields on the boundary
 of a quantum Hall droplet, thereby eliminating CS type (potentially anomaly-generating)
 terms in the bulk. We showed that this does lead to the same results
 in 2+1 dimensions, same as integrating out the $b$-fields and
as obtained in various explicit calculations.
 Anomaly cancellation thus constitutes an alternate formulation of the
 key idea of the parton picture and this is indeed generalizable to
 higher dimensions.
 
We worked out the parton picture in 4+1 dimensions in section 4.
The effective action for the partons
can be obtained as the Chern-Simons term
corresponding to the twisted Dolbeault index density
 in $2k+2$ dimensions, for QHE in $2k+1$ dimensions.
 The reason for the use of this index is the same as for
 the integer QHE, namely, because of the holomorphicity
 condition for the fields in the lowest Landau level.
 The spectator fields are chiral spinors. Their anomaly can be 
 obtained by the standard descent procedure
 starting with the Dirac index density in
 $2k+ 2$ dimensions.
The resulting effective actions are given in
(\ref{4d-11}) and (\ref{122}).

The Hall current is given by deriving this action with respect to the
electromagnetic field. Likewise, the variation of the action
with respect to the metric gives the energy-momentum tensor.
From the current and the energy-momentum tensor, we obtained the
Hall conductivity and the Hall viscosity, given in
(\ref{4d-13}) and (\ref{140}).
The $m$-dependence for the
leading term for the Hall conductivity is $1/m^2$ while it is
$1/m$ for Hall viscosity. 

A related point worthy of comment is about the case of QHE on $S^4$ studied by Hu and Zhang in \cite{ZH}. The authors argue that the filling fraction for the corresponding Laughlin-type wavefunctions is $1/m^3$, based on the analysis of the degeneracy and the spectrum of the lowest Landau level.\footnote{ We thank the referee for pointing this out.} While this is different from the $1/m^2$ behavior of the Hall current for a complex 4d-manifold, it is easy to see how this arises.
We have shown in \cite{KN1} that QHE on $S^4$ with $SU(2)$ magnetic background can be understood
in terms of QHE on $\mathbb{CP}^3$ with an abelian magnetic field. This arises from the fact that $\mathbb{CP}^3$ is an $S^2$-bundle over 
$S^4$. The filling fraction of the corresponding Laughlin-type wave functions on $\mathbb{CP}^3$ is also $\nu = 1 /m^3$. In terms of the effective action approach, 
the leading term of the CS action for the electromagnetic field
$A$ will be proportional to $1/m^3$ for $\mathbb{CP}^3$;
this is in accordance with \cite{ZH}.

In fact we expect that for a general complex manifold of dimension $2k$ for which we can construct an effective action for Laughlin type states as described in this paper, the leading term for the Hall conductivity will be proportional to $\nu ={1 \over m^k}$ while the Hall viscosity will scale as $1 \over m^{k-1}$, up to curvature corrections. This is because the dominant term in the effective action for the Hall current is the CS term $\int A (dA/m)^k$, while the dominant term for the derivation of the Hall conductivity is the next order term of the form $\int A (dA/m)^{k-1} R$, involving one power of the curvature and hence one less power of $A$.

We have only considered states which are the higher dimensional analogs of the Laughlin states, i.e., of the $\nu = 1/m^k$ type.
For other fractional values of $\nu$,
ensuring that each species of partons has the same degeneracy
(over the integer Landau levels they fill) is nontrivial even in
2+1 dimensions for spaces of nontrivial topology, as argued at the end of section 3.

Another point worth emphasizing is about the
use of the spectator fields.
It is important that the picture
of the electron as a composite particle made of
partons is to be viewed only as a theoretical technique
highlighting certain nonperturbative features of 
the mutually interacting electrons in a magnetic field.
While it is useful and seems to work well in 2+1 dimensions, the ultimate
reason for its success is still unclear.
In reality, the only physical particle involved is just the electron.
So while response functions derivable from the
effective action are to be viewed as physical, 
possible composites of the spectator fields are to be viewed as
artifacts of the technique. For this reason, we do not think the excitations
of $\Xi$  in (\ref{fqhe24}), or similar fields in 4+1 dimensions,
 are to be viewed as physical.
 
 Obviously, the generalization of the present work to 
 arbitrary even dimensions
 and to nonabelian gauge field backgrounds
 will be very interesting. In envisaging such prospects, we note that
 the issue of anomaly cancellation becomes more involved. 
In 4+1 dimensions, in (\ref{4d-9}),
we encountered a term corresponding to
a mixed gauge-gravitational anomaly.
In even higher dimensions, there are additional terms corresponding to
mixed gauge-gravitational anomalies generated by the index density.
A consistent anomaly cancellation scenario incorporating these elements
is beyond the scope of this first attempt, but nevertheless 
it remains a worthwhile avenue to explore.
 
\bigskip

We thank Hans Hansson for a careful reading of the manuscript and
useful comments.
Part of this work was done during a recent visit by two of us
(DK, VPN) to the Dublin Institute for Advanced Studies (DIAS), Dublin, Ireland.
We thank Denjoe O'Connor and members
of the School of Theoretical Physics at DIAS for their warm hospitality. 
We also acknowledge hospitality at the Indian Institute of Science, Bangalore (AA, VPN) and at the Raman Research Institute, Bangalore (DK) during earlier research visits there.

This work was supported in part by the U.S. National Science Foundation Grants No. PHY-2112729, PHY-2412479 and PHY-2412480, and by a PSC-CUNY grant.



\begin{thebibliography}{99}

\bibitem{QHE-general} 
Since QHE is an old topic with a vast literature, we refer 
to recent reviews and books: 
R.E. Prange and S.M. Girvin, {\it The Quantum Hall Effect},
2nd ed. (Springer-Verlag, Berlin, 2012); Z.F. Ezawa,
{\it Quantum Hall Effects} (World Scientific, Singapore, 2008);
T.H. Hansson, M. Hermanns, S.H. Simon and S.F. Viefers, \RMP~{\bf 89}, 025005 (2017);
D. Tong, {\it Lectures on quantum Hall effect}, arXiv:1606.06687[hep-th].

\bibitem{4DQHE} Y. E. Kraus, Z. Ringel, and O. Zilberberg, 
Phys. Rev. Lett. {\bf 111}, 226401 (2013); H.M. Price, O. Zilberberg, T. Ozawa, I. Carusoto and N. Goldman
\PRL ~{\bf 115}, 195303 (2015); 
\PR~{\bf B 93}, 245113 (2016); T. Ozawa, H. M. Price, N. Goldman, O. Zilberberg, and I. Carusotto, 
Phys. Rev. {\bf A93}, 043827 (2016); 
O. Zilberberg, S. Huang,
J. Guglielmon, M. Wang, K. P. Chen, Y. E. Kraus, and M. C. Rechtsman, 
Nature {\bf 553}, 59 (2018);
M. Lohse, C. Schweizer, H.M. Price, O. Zilberberg, and I. Bloch, 
Nature {\bf 553}, 55 (2018).

\bibitem{4DQHE1} A. Fabre, J.B. Bouhiron, T. Satoor, R. Lopes and S. Nascimbene, 
arXiv:2210.06322. 

\bibitem{ZH} S.C. Zhang and J.P. Hu, 
 {Science} {\bf 294} (2001) 823; 
J.P. Hu and S.C. Zhang, 
\PR~{\bf B 66}, 125301 (2002);
see also
J. Fr\"ohlich and U.M. Studer, 
Commun. Math. Phys. {\bf 148}, 553 (1992);
\RMP { \bf 65}, 733 (1993).

\bibitem{KN1} D. Karabali and V.P. Nair, 
\NP { \bf B 641}, 533 (2002);
 \NP { \bf B 679}, 427 (2004);
 \NP { \bf B 697}, 513 (2004).
 
\bibitem{bosonization} D. Karabali, Nucl. Phys. {\bf B 726}, 407 (2005); Nucl. Phys. {\bf B 750}, 265 (2006); V.P. Nair, Nucl.
Phys. {\bf B 750}, 289 (2006).

\bibitem{KN4} D. Karabali, V.P. Nair and S. Randjbar-Daemi,  {\it Fuzzy spaces, the M(atrix) model and quantum Hall effect}, published in  {\it From fields to strings: Circumnavigating theoretical physics}, edited by M. Shifman et al., vol. 1, 831-875 (World Scientific, Singapore, 2005); D. Karabali and V.P. Nair, J. Phys A: Math. Gen. {\bf 39}, 12735 (2006).


 \bibitem{alexios} A.P. Polychronakos, Nucl. Phys.
{\bf B 705}, 457 (2005); Nucl. Phys. {\bf B 711}, 505 (2005).

\bibitem{everyone} 
H. Elvang and J. Polchinski, 
C.R. Physique {\bf 4}, 405 (2003); 
B.A. Bernevig, C.H. Chern, J.P. Hu, N. Toumbas and S.C. Zhang, 
{ Ann. Phys.} {\bf 300}, 185 (2002);
B. A. Bernevig, J.P. Hu, N. Toumbas and S.C. Zhang, 
{\bf 91}, 236803 (2003); 
 G. Meng, 
 { J. Phys.} {\bf A36}, 9415 (2003);  
 V.P. Nair and S. Randjbar-Daemi, 
 { Nucl. Phys.} {\bf B679}, 447 (2004); 
 A. Jellal, 
 { Nucl. Phys.} {\bf B725}, 554 (2005);  K. Hasebe, 
 Nucl. Phys. {\bf B886}, 952 (2014).

 \bibitem{VPN-Rand} V.P. Nair and S. Randjbar-Daemi, 
 { Nucl. Phys.} {\bf B679}, 447 (2004).
 
\bibitem{EGH} T. Eguchi, P.B. Gilkey and A.J. Hanson, {\it Gravitation, Gauge Theories and Differential Geometry}, Phys. Rep. { \bf 66}, 213 (1980).

\bibitem{KN5} D. Karabali and V.P. Nair, \PR~{\bf D 94}, 
024022 (2016)\\
{}[arXiv:1604.00722[hep-th]].


\bibitem{WZ} X. Wen and A. Zee,
  \PRL { \bf 69}, 953 (1992).
  
 \bibitem{visc}
J.E. Avron, R. Seiler and P.G. Zograf,
 \PRL~ { \bf 75}, 697 (1995);
N. Read, 
\PR~ {\bf B 79}, 045308 (2009); N. Read and E.H. Rezayi, 
\PR~ { \bf B 84}, 085316 (2011);
C. Hoyos and D.T. Son,
 \PRL ~{\bf 108}, 066805 (2012). 


\bibitem{AG1} A.G. Abanov and A. Gromov,
 \PR  { \bf B90}, 014435 (2014); 
 A. Gromov and A. G. Abanov, 
  \PRL { \bf 113}, 266802 (2014).
  
 \bibitem{framing} A. Gromov, G. Cho, Y. You, A.G. Abanov and E. Fradkin, 
{ \PRL} { \bf 114}, 016805 (2015).

\bibitem{AG2} 
T. Can, M. Laskin and P. Wiegmann, 
\PRL { \bf 113}, 046803 (2014); 
Ann. Phys. { \bf 362} 752 (2015);
S. Klevtsov and P. Wiegmann,
  \PRL { \bf 115} 086801 (2015);
B. Bradlyn and N. Read, 
\PR { \bf B91}, 165306 (2015); 
S. Klevtsov, X. Ma, G. Marinescu and P. Wiegmann, 
Commun. Math. Phys. {\bf 349}, 819 (2017).

\bibitem{KN6} D. Karabali and V.P. Nair, \PR~{\bf B 108}, 205155 (2023)
[arXiv:2307.15919].

\bibitem{books}
J. K. Jain, {\it Composite Fermions},
(Cambridge University Press, 2007);
R. Shankar,
{\it Quantum Field Theory and Condensed Matter},
(Cambridge University Press, 2017).

\bibitem{Gromov-thesis}
A. Gromov, {\it Geometric Aspects of Quantum Hall States},
(Doctoral dissertation, The Graduate School, Stony Brook University, 2015). 

\bibitem{AA-FQH}
A.~Agarwal,
J. Phys. A \textbf{55}, no.2, 025402 (2022);
[arXiv:2106.10793 [hep-th]].

\bibitem{partons1} J.K. Jain, \PR~{\bf B 40}, 8079 (1989); \PR~{\bf B 41}, 7653 (1990).

\bibitem{partons2} X.G. Wen, \PRL~{\bf 66}, 802 (1991); \PR~{\bf B 60}, 8827 (1999).

\bibitem{heck} J.J. Heckman and L. Tizzano, JHEP 05 (2018) 120;
G. Palumbo, JHEP 05 (2022) 124.
An earlier work on fractional QHE in higher dimensions using higher rank differential forms and related flux attachment techniques is K. Hasebe, \NP~{\bf B886}, 952 (2014).

\bibitem{anom} See, for example,
R.A. Bertlmann, {\it Anomalies in Quantum Field Theory} (Oxford University Press, New York, 1996);
S. Treiman, R. Jackiw, B. Zumino and E. Witten,
{\it Current Algebra and Anomalies}  (World Scientific, Singapore, 1985).

\bibitem{gilkey} P.B. Gilkey, {\it Invariance Theory,
the Heat Equation and the Atiyah-Singer Index Theorem},
Mathematics Lecture Series Vol. 11, Publish or Perish Inc. (1984),
ISBN 0-914098-20-9.

\end{thebibliography}
\end{document}